\documentclass[conference,a4paper]{IEEEtran}

\usepackage[T1]{fontenc}
\usepackage{graphicx}

\usepackage{amsthm,amsmath,amssymb,bm}
\usepackage{mathrsfs}
\usepackage{amsfonts}
\newtheorem{theorem}{Theorem}

\newtheorem{lemma}{Lemma}

\newtheorem{remark}{Remark}

\newtheorem{proposition}{Proposition}
\usepackage{subfigure}
\usepackage{algorithm}
\usepackage{algpseudocode}
\usepackage{graphics}
\usepackage{epsfig}
\usepackage{cite}
\usepackage{color}
\usepackage{booktabs}
\usepackage{verbatim}
\usepackage[colorlinks,linkcolor=blue,urlcolor=blue,citecolor=blue]{hyperref}

\begin{document}
\title{Model-Driven Deep Learning for Distributed Detection with Binary Quantization}
\author{
	\IEEEauthorblockN{Wei Guo\IEEEauthorrefmark{1}, Meng He\IEEEauthorrefmark{2}, Chuan Huang\IEEEauthorrefmark{2}, Hengtao He\IEEEauthorrefmark{1}, Shenghui Song\IEEEauthorrefmark{1}, Jun Zhang\IEEEauthorrefmark{1}, and Khaled B. Letaief\IEEEauthorrefmark{1}}	
	\IEEEauthorblockA{\IEEEauthorrefmark{1}Dept. of Electronic and Computer Engineering, Hong Kong University of Science and Technology, Hong Kong SAR.}
	\IEEEauthorblockA{\IEEEauthorrefmark{2}School of Science and Engineering and Future Network of Intelligence Institute,\\ 
		The Chinese University of Hong Kong, Shenzhen 518172, China.}
	\IEEEauthorblockA{Emails: \IEEEauthorrefmark{1}\{eeweiguo, eehthe, eeshsong, eejzhang, eekhaled\}@ust.hk, \IEEEauthorrefmark{2}menghe@link.cuhk.edu.cn, \IEEEauthorrefmark{2}huangchuan@cuhk.edu.cn}
}

\maketitle

\begin{abstract}
Within the realm of rapidly advancing wireless sensor networks (WSNs), distributed detection assumes a significant role in various practical applications. However, critical challenge lies in maintaining robust detection performance while operating within the constraints of limited bandwidth and energy resources. This paper introduces a novel approach that combines model-driven deep learning (DL) with binary quantization to strike a balance between communication overhead and detection performance in WSNs. We begin by establishing the lower bound of detection error probability for distributed detection using the maximum a posteriori (MAP) criterion. Furthermore, we prove the global optimality of employing identical local quantizers across sensors, thereby maximizing the corresponding Chernoff information. Subsequently, the paper derives the minimum MAP detection error probability (MAPDEP) by inplementing identical binary probabilistic quantizers across the sensors. Moreover, the paper establishes the equivalence between utilizing all quantized data and their average as input to the detector at the fusion center (FC). In particular, we derive the Kullback-Leibler (KL) divergence, which measures the difference between the true posterior probability and output of the proposed detector. Leveraging the MAPDEP and KL divergence as loss functions, the paper proposes model-driven DL method to separately train the probability controller module in the quantizer and the detector module at the FC. Numerical results validate the convergence and effectiveness of the proposed method, which achieves near-optimal performance with reduced complexity for Gaussian hypothesis testing.
\end{abstract}


\IEEEpeerreviewmaketitle

\section{Introduction}\label{sec:intro}
The recent advancements of wireless sensor networks (WSNs) have elevated its significance as the underlying infrastructure for wide range of practical applications. These applications encompass diverse fields, including environmental monitoring, military surveillance \cite{chong2003sensor}, smart industries, and smart grids driven by 5G/6G networks \cite{gururaj2023collaborative,letaief2019roadmap}. To this end, the detection of events of interest by distributed sensors, usually performed through binary hypothesis testing, is regarded as the fundamental problem and has been investigated thoroughly. A typical WSN consists of multiple sensors distributed at different locations and a fusion center (FC) aggregates the local observations to make the final decision. In practice, due to the bandwidth and energy limitations, the local observations need to be quantized into finite number of bits before transmission to the FC, which inevitably incurs degradation of detection performance \cite{viswanathan1997distributed}.

To balance the tradeoff between the communication overhead and detection performance degradation, a bunch of approaches have been proposed \cite{tenney1981detection,zhang2002optimal, altay2016optimal, gul2021fast,gul2021scalable}. In \cite{tenney1981detection}, distributed detection with binary quantization level was studied without considering the FC design based on the Bayesian statistics. Then, the authors of \cite{zhang2002optimal} determined the optimal fusion rule for both the Bayesian and Neyman-Pearson detection problem with binary quantization. Based on the Chernoff information and deflection criterion, the authors in \cite{altay2016optimal} found the optimal local quantization intervals for independent and identically distributed (i.i.d.) sensors. In the most recent work of \cite{gul2021fast,gul2021scalable}, the authors proposed a Gaussian approximation method for fast and scalable multilevel quantization and FC design with both i.i.d. and non-i.i.d. sensors. However, current state-of-the-art approaches for distributed detection assume perfect knowledge about the prior probabilities of the observation noise and distributions of the hypothesis, which are generally not valid in reality. 

To solve the problem, this paper proposes a model-driven deep learning (DL) method \cite{he2020model} for the distributed detection problem in WSNs with binary quantization. Specifically, we focus on distributed hypothesis testing and consider the case that the sensors utilize binary quantizers with conditionally i.i.d. observation noise. We first derive the minimum achievable detection error probability based on the maximum a posteriori (MAP) criterion, which can serve as the performance metric for quantizer design. Then, we prove the global optimality of utilizing identical quantizer by maximizing the Chernoff information corresponding to the MAP detection error probability (MAPDEP). Based on the above results, we utilize the identical binary probabilistic quantizer across the sensors and derive the corresponding MAPDEP as the objective to optimize the probability controller in the quantizer. Next, we prove that employing the mean value of the local quantized data for detection at the FC achieves global optimal performance and derive the Kullback-Leibler (KL) divergence between the true posterior probability and proposed detector. Finally, based on the model-driven DL principle, we propose a separate training method to successively train the probability controller at the sensors and detector modules at the FC by utilizing the MAPDEP and KL divergence as the loss functions, respectively. Simulation results demonstrate the proposed method has similar performance with the optimal one but with reduced complexity for Gaussian hypothesis testing. 

\section{System Model and Problem Formulation}\label{sec:model_problem}
\subsection{System Model}\label{subsec:system_model}
\begin{figure}[!htbp]
	\centering
	\includegraphics[width=0.45\textwidth]{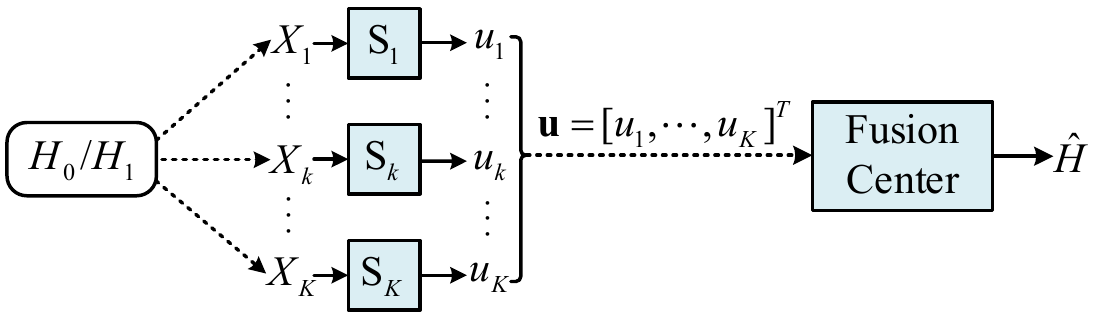}
	\caption{System Model.}
	\label{fig:system}
\end{figure}

As shown in Fig. \ref{fig:system}, we consider the distributed detection in WSN, where the FC aims to detect the desired state of nature based on the observations received from $K$ distributed sensors, denoted as $\mathrm{S}_k,\ k=1,\cdots, K$. Specifically, we consider the distributed hypothesis testing problem, where the true hypothesis $H$ belongs to the binary set $\{H_0,H_1\}$ with prior probabilities $p(H=H_0)=\pi_0$ and $p(H=H_1)=\pi_1$. Each sensor $\mathrm{S}_k, k=1,\cdots,K,$ independently observes $H$ and obtains the local observation $X_k$, which is a noisy version of $H$. 

We consider the scenario where the observation noise at all sensors is i.i.d., and thus the local observations obtained at the sensors are conditionally i.i.d. with given $H$, i.e., 
\begin{equation}\label{eq:local_noise_distribution}
	\begin{aligned}
		&f_{X_1, \cdots, X_K \mid H}\left(x_1, \cdots, x_K \mid H_n\right)\\
		&=\prod_{k=1}^K f_{X_k \mid H}\left(x_k \mid H_n\right)=\prod_{k=1}^K f_{X \mid H}\left(x_k \mid H_n\right), n=0,1,
	\end{aligned}
\end{equation} 
where $\forall x_1,\cdots,x_K\in\mathbb{R}$, $f_{X_1, \cdots, X_K \mid H}\left(\cdot \mid \cdot\right)$ is the conditional joint probability density function (PDF) of all observations with original state $H$, and $f_{X \mid H}(\cdot \mid \cdot)=f_{X_1 \mid H}(\cdot \mid \cdot)=\cdots=f_{X_K \mid H}(\cdot \mid \cdot)$ is the conditional marginal PDF of the local observation with given $H$ at each sensor.

Due to the bandwidth limitations, each sensor $\mathrm{S}_k$, $k=1,\cdots,K$, needs to first quantize its respective local observation $X_{k}$ into a discrete message $u_k$, whose quantization level is $L$, before transmitting the signal to FC.
Let the conditional distribution $p(u_k|X_k)$ describes the probabilistic quantization process at sensor $k$, specifying the probability distribution of the quantized data $u_k$ given the local observation $x_k$.
Subsequently, each sensor $\mathrm{S}_k$, $k=1,\cdots,K$, transmits $u_k$ to the FC through an error free channel, and the FC aggregates the quantized data  $\mathbf{u}=[u_1,\cdots,u_K]^T$ from all $K$ sensors to generate a recovered state, denoted as $\hat{H}$, which represents an hypothesis of the true state.
\subsection{Problem Formulation}\label{subsec:problem}
To evaluate the detection performance, we utilize the probability of detection error $P_e(H,\mathbf{u},\hat{H})$ as the cost function. That is
\begin{align}\label{eq:detection_error}
	P_e(H,\mathbf{u},\hat{H})=&p\left(\hat{H}\neq H\right)\nonumber\\
	=&\pi_0p\left(\hat{H}=H_1|H_0\right)+\pi_1p\left(\hat{H}(\mathbf{u})=H_0|H_1\right)\nonumber\\
	=&\pi_{0}\sum_{\mathbf{u}}
	p\left(\hat{H}(\mathbf{u})=H_{1}|\mathbf{u}\right)p\left(\mathbf{u}|H_{0})\right)\nonumber\\
	&+
	\pi_{1}\sum_{\mathbf{u}}
	p\left(\hat{H}(\mathbf{u})=H_{0}|\mathbf{u}\right)p(\mathbf{u}|H_{1}).
\end{align}
Our goal is to find the optimal quantizers $\{p(u_k|X_k)\}_{k=1}^K$ and detector $\hat{H}(\mathbf{u})$ to minimize $P_e(H,\mathbf{u},\hat{H})$, and the corresponding optimization problem is formulated as
\begin{equation}\label{eq:opt_problem}
	\underset{\left\{p(u_k|X_k)\right\}_{k=1}^{K},\,
		\hat{H}(\cdot)}{\min}\
	P_e(H,\mathbf{u},\hat{H}).
\end{equation}

\section{Binary Quantization and Detector Design}\label{sec:quantizer_FC_design}
In this section, we consider the design of binary quantization at the sensors and the detector at the FC to minimize $P_e(H,\mathbf{u},\hat{H})$ in problem \eqref{eq:opt_problem}. Since we cannot directly derive the closed-form expression of $P_e(H,\mathbf{u},\hat{H})$, we first find the attainable performance bound of $P_e(H,\mathbf{u},\hat{H})$. To this end, we deploy the MAP detector at the FC, i.e.,
\begin{equation}\label{eq:MAP_rule}
	\hat{H}_{\mathrm{MAP}}(\mathbf{u})=\begin{cases}
		H_0,\ \dfrac{\pi_0p(\mathbf{u}|H_0)}{\pi_1p(\mathbf{u}|H_1)} >1,\\
		H_1,\ \dfrac{\pi_0p(\mathbf{u}|H_0)}{\pi_1p(\mathbf{u}|H_1)} \leq1.
	\end{cases}
\end{equation} 
Then, we can obtain the minimum achievable detection error probability with the MAP detector, denoted as MAPDEP, in the following lemma.

\begin{lemma}\label{lem:MAPDEP}
	The MAP decision rule minimizes the probability of detection error at the FC \cite{cover2006elements}, with the minimum achievable detection error probability given as
	\begin{equation}\label{eq:MAPDEP}
		P_e(H,\mathbf{u},\hat{H})
		\geq P_{E}(H,\mathbf{u})
		\triangleq\sum_{\mathbf{u}}\min\left\{\pi_0p(\mathbf{u}|H_0),\pi_1p(\mathbf{u}|H_1)\right\}.
	\end{equation}
\end{lemma}
\begin{IEEEproof}
	Since the MAP detector \eqref{eq:MAP_rule} is deployed at the FC, we have
	\begin{equation}\label{eq: MAPDEP_proof}
		\begin{aligned}
			P_e(H,\mathbf{u},\hat{H})
			\geq&
			P_e(H,\mathbf{u},\hat{H}_{\mathrm{MAP}})\\
			=&\pi_{0}\sum_{\mathbf{u}}
			p(\hat{H}_{\mathrm{MAP}}(\mathbf{u})=H_{1}|\mathbf{u})p(\mathbf{u}|H_{0})\\
			&+\pi_{1}\sum_{\mathbf{u}}
			p(\hat{H}_{\mathrm{MAP}}(\mathbf{u})=H_{0}|\mathbf{u})p(\mathbf{u}|H_{1})\\
			=&\sum_{\mathbf{u}:\pi_0p(\mathbf{u}|H_0)>\pi_1p(\mathbf{u}|H_1)}
			\pi_{0}p(\mathbf{u}|H_{0})\\
			&+
			\sum_{\mathbf{u}:\pi_0p(\mathbf{u}|H_0)\leq\pi_1p(\mathbf{u}|H_1)}
			\pi_{1}p(\mathbf{u}|H_{1})\\
			=&\sum_{\mathbf{u}}\min\left\{\pi_0p(\mathbf{u}|H_0),\pi_1p(\mathbf{u}|H_1)\right\}.
		\end{aligned}
	\end{equation}
	Define $P_{E}(H,\mathbf{u})=P_e(H,\mathbf{u},\hat{H})$ as MAPDEP, which completes the proof.
\end{IEEEproof}
\begin{remark}\label{rem:MAPDEP_metric}
	The expression of $P_{E}(H,\mathbf{u})$ in \eqref{eq:MAPDEP} is determined only by the prior probability $\pi_n,n=0,1$, and the quantized data $\mathbf{u}$, which means MAPDEP $P_{E}(H,\mathbf{u})$ can serve as a performance metric for quantizer design.  
\end{remark}

There is another closely related metric to MAPDEP in hypothesis testing, i.e., Chernoff information \cite{cover2006elements}. According to the reciprocity between the MAPDEP and Chernoff information, i.e., higher Chernoff information implies smaller MAPDEP and vice versa, minimizing $P_{E}(H,\mathbf{u})$ is equivalent to maximizing the Chernoff information. As a result, we will use Chernoff information as our performance metric to prove the global optimality of identical quantizer.

\begin{lemma}\label{lem:glob_opt_identical_quantizer}
	If all sensors have conditionally i.i.d. local observations as shown in \eqref{eq:local_noise_distribution} and the identical quantization level $L$, using identical quantizer design at all sensors is globally optimal in terms of the Chernoff information maximization, and the maximum average Chernoff information is given as 
	\begin{equation}\label{eq:max_Chernoff}
		C(H,\mathbf{u})=-\underset{0\leq \alpha\leq 1}{\min} \log\left[\sum_{u=0}^{L-1}p(u|H_{0})^{\alpha}p(u|H_{1})^{1-\alpha}\right].
	\end{equation}  
\end{lemma}
\begin{IEEEproof}
	Please see Appendix. \ref{append:proof_identical_quantizer}. 
\end{IEEEproof}

Since all sensors deploy an identical quantizer, the original problem \eqref{eq:opt_problem} can be further simplified as  
\begin{equation}\label{eq:opt_problem_quantizer}
	\underset{p(u|X)}{\min}\
	P_E(H,\mathbf{u}),
\end{equation}
which helps us to investigate the quantizer and detector design.
\subsection{Binary Quantizer Design}\label{subsec:binary_quantizer}
In this paper, we consider the binary quantization scheme. According to Lemma \ref{lem:glob_opt_identical_quantizer}, we consider the case that all sensors adopt an identical binary quantizer, and the following proposition shows the minimum achievable MAPDEP in this case.
\begin{proposition}\label{prop:MAPDEP_binary_quantize}
	If the binary quantized data $u_1,\cdots,u_K$ from all sensors are conditionally i.i.d. with given $H$, then the minimum achievable MAPDEP of $H$ is given as 
	\begin{equation}\label{eq:MAPDEP_Binary_Quantization}
		\begin{aligned}
			P_E^{\text{binary}}\triangleq\sum_{k=0}^{K}C_{K}^{k}\min&\left\{\pi_0(\gamma(H_0))^k(1-\gamma(H_0))^{K-k},\right.\\
			&\left.\pi_1(\gamma(H_1))^kp(1-\gamma(H_1))^{K-k}\right\},
		\end{aligned}
	\end{equation}
	where 
	\begin{equation}\label{eq: gamma(H) 1}
		\gamma(H)=p(u=1|H)=\mathbb{E}_{X}[p(u=1|X)|H],\ H\in\{H_0,H_1\},
	\end{equation}
	denotes the conditional probability of any quantized data being `1' with given $H$ and $C_K^k$ is the binomial coefficient.
\end{proposition}
\begin{IEEEproof}
	Please see Appendix. \ref{append:proof_MAPDEP_binary_quantize}.  	 	
\end{IEEEproof}

\begin{figure}[!htbp]
	\centering
	\includegraphics[width=0.45\textwidth]{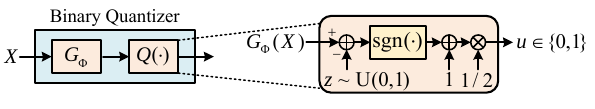}
	\caption{Proposed Binary Quantizer.}
	\label{fig:quantizer}\vspace{-1em}
\end{figure}

The expression of $P_E^{\text{binary}}$ in \eqref{eq:MAPDEP_Binary_Quantization} is valid for all kinds of binary quantizer design. As shown in Fig. \ref{fig:quantizer}, we consider the binary probabilistic quantizer with random dithering \cite{gray1993dithered}. Specifically, the local observation $X$ is first sent to a probability controller $G(\cdot): \mathbb{R}\rightarrow [0,1]$, and then the output $G(X)$ is fed into a quantization function $Q(\cdot):[0,1]\rightarrow\{0,1\}$, to generate a random binary data $u$, which is given as
\begin{equation}\label{eq:quantization_func}
	u=Q(G(X))=\frac{1+\text{sgn}(G(X)-z)}{2}\in\{0,1\}.
\end{equation} 
Here, $z\sim U(0,1)$ is a standard uniform distributed dithering noise and $\text{sgn}(\cdot)$ is the sign function. As shown in \cite{he2024joint}, such quantizer ensures that $p(u=1|X)=G(X)$ and $p(u=0|X)=1-G(X)$. 

Our goal is to find the optimal binary quantizer to minimize $P_E^{\text{binary}}$ in \eqref{eq:MAPDEP_Binary_Quantization}, and the key to find the optimal probability controller $G(\cdot)$. However, $P_E^{\text{binary}}$ is determined by the conditionally probability of quantized data from all sensors, which is hard to obtain in reality. This motivates us to use model-driven DL method \cite{he2020model} to train a deep neural network (DNN), parameterized as $\Phi\in\mathbb{R}^d$, to approximate the optimal probability controller $G^*(\cdot)$. The details will be presented in Section \ref{sec:DL_Opt}.\vspace{-1em}

\subsection{Detector Design}\label{sec:detector_design}
\begin{figure}[!htbp]
	\centering
	\includegraphics[width=0.45\textwidth]{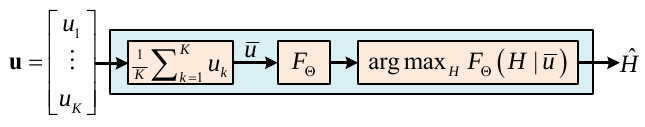}
	\caption{Proposed Detector.}
	\label{fig:fc}\vspace{-1em}
\end{figure}

After obtaining the DNN-based binary quantizer, we further need to find the optimal detector at the FC to minimize $P_e(H,\mathbf{u},\hat{H})$ in problem \eqref{eq:opt_problem}. As shown in Section \ref{subsec:system_model}, the FC aggregates the quantized data $\mathbf{u}\in\{0,1\}^K$ from all users to generate the final decision $\hat{H}$. Based on the MAP decision rule, we prove the equivalence between utilizing all the quantized data and their average as the input to the detector at the FC in the following theorem.
\begin{theorem}\label{th:opt_mean_fusion}
	If the binary quantized data $u_k$, $k=1,\cdots,K$, from all sensors are conditionally i.i.d. with given $H$, the detection of $H$ by using the quantized data $\mathbf{u}=[u_1,\dots,u_K]^T$ and their average $\bar{u} = \frac{1}{K}\sum_{k=1}^{K} u_k$ achieve the identical minimum detection error probability, i.e.,
	\begin{equation}\label{eq:mean_fusion_MAPDEP}
		\begin{aligned}
			P_{E}(H,\mathbf{u})=&P_E^{\text{binary}}\\
			=&P_{E}(H,\bar{u})\triangleq\sum_{\bar{u}}\min\left\{\pi_0p(\bar{u}|H_0),\pi_1p(\bar{u}|H_1)\right\},
		\end{aligned}
	\end{equation}
	where $P_{E}(H,\bar{u})$ is the minimum detection error probability achieved by using $\bar{u}$ for the detection of $H$ and $P_E^{\text{binary}}$.
\end{theorem}
\begin{IEEEproof}
	Please see Appendix. \ref{append:proof_mean_fusion}.
\end{IEEEproof}
As illustrated in Fig. \ref{fig:fc}, we can use the average $\bar{u}$ instead of all the quantized data $\mathbf{u}$ as the input for the detector. Then, the MAP decision for the detection of $H$ is given by
\begin{equation}\label{eq:MAP_detector_average}
	\hat{H}_{\text{MAP}}(\bar{u})=\arg\max_{H}\ p(H|\bar{u}).
\end{equation}
However, the posterior probability $p(H|\bar{u})$ is also very difficult to be obtained in reality. Similarly, we can also use a well-trained DNN, parameterized as $\Theta$, to approximate the true posterior probability. The details will be presented in Section \ref{sec:DL_Opt}. \vspace{-1em}

\section{Model-Driven Deep Learning for Quantizer and Detector Optimization}\label{sec:DL_Opt}
As shown in Section \ref{sec:quantizer_FC_design}, the DNN $\Phi$ is trained to approximate the optimal probability controller $G^*(\cdot)$, and thus the dimensions of the input and output of $\Phi$ are both 1. Since we already obtained a metric, i.e., the minimum achievable MADPDEP $P_E^{\text{binary}}$, which can evaluate the performance of the binary quantizer. Thus, it is quite straightforward that we can use $P_E^{\text{binary}}$ as the loss function to evaluate the training performance of $\Phi$. Let $G_{\Phi}(\cdot)$ denotes the probability controller parameterized by $\Phi$, then the loss function for training $\Phi$ is given by 
\begin{equation}\label{eq:loss_func_quantizer}
	\begin{aligned}
		\mathcal{L}_{\Phi}=\sum_{k=0}^{K}C_{K}^{k}\min&\{\pi_0(\gamma_{\Phi}(H_0))^kp(1-\gamma_{\Phi}(H_0))^{K-k},\\
		&\pi_1(\gamma_{\Phi}(H_1))^kp(1-\gamma_{\Phi}(H_1))^{K-k}\},
	\end{aligned}
\end{equation}
with $\gamma_{\Phi}(H_n)=\mathbb{E}_{X}[G_{\Phi}(X)|H_n]=\int_x f_{X|H}(x|H_n)\ dx,\ n=0,1$.

As for the training of the DNN $\Theta$ to approximate the true posterior probability, it is hard to find a existing tractable metric for loss function. Let $F_{\Theta}(\cdot):[0,1]\rightarrow[0,1]\times[0,1]$ be the posterior approximator parameterized by $\Theta$, and the output is  $F_{\Theta}(H\mid\bar{u})=\left[F_{\Theta}(H_0\mid\bar{u}), F_{\Theta}(H_1\mid\bar{u})\right]$ with given $\bar{u}$. It is intuitively to see that the closer between the true posterior probability $p(H|\bar{u})$ and $F_{\Theta}(\bar{u})$, the better detection performance is achieved by the proposed detector. Therefore, we utilize the average KL divergence \cite{cover2006elements}, i.e., $\mathbb{E}_{\bar{u}}\left[D_{KL}(p(H|\bar{u})|F_{\Theta}(H\mid\bar{u}))\right]$, to describe their similarity, which can serve as the performance metric for the proposed detector.
\begin{proposition}\label{prop:KL_diverngence}
	With the proposed binary quantizer and detector design, the average KL divergence between $p(H|\bar{u})$ and $F_{\Theta}(H\mid\bar{u})$ is
	\begin{equation}\label{eq:KL_divergence}
		\begin{aligned}
			&D_{KL}^{\text{binary}}(\Phi,\Theta)\\
			=&\ \mathbb{E}_{\bar{u}}\left[D_{KL}(p(H|\bar{u})|F_{\Theta}(H\mid\bar{u}))|\Phi,\Theta\right]\\
			=&\
			\sum_{n=0}^{1}\sum_{k=0}^{K}
			\pi_np_{\Phi}^{k}(H_n)
			\log\frac{\pi_np_{\Phi}^{k}(H_n)}{F_{\Theta}(H_n|\bar{u})\sum_{n=0}^{1}\pi_np_{\Phi}^{k}(H_n)},
		\end{aligned}
	\end{equation}
	where $p_{\Phi}^{k}(H_n)=C_K^k(\gamma_{\Phi}(H_n))^k(1-\gamma_{\Phi}(H_n))^{K-k}$.
\end{proposition}
\begin{IEEEproof}
	Please see Appendix. \ref{append:proof_KL_divergence}.
\end{IEEEproof}
As mentioned before, the quantizer, equivalently the probability controller, is independent of the detector with the MAP decision rule. It can be achieved by training a DNN $\Phi$ with the loss function $\mathcal{L}_{\Phi}$ derived in \eqref{eq:loss_func_quantizer}. After obtaining the DNN parameters $\Phi^*$, we can derive the average KL divergence \eqref{eq:KL_divergence}. Furthermore, the loss function for training $\Theta$ can be defined as $\mathcal{L}_{\Theta}=D_{KL}^{\text{binary}}(\Phi^*,\Theta)$. Thus, a separate training method is proposed to obtain the DNN parameters $\Phi^*$ and detector DNN parameters $\Theta^*$.

We consider the separate training of $\Phi$ and $\Theta$ under the scenario where the observation noise is unknown. The training process is based on the dataset $D_1=\{x_{t,0},x_{t,1}\}_{t=1}^T$, where $x_{t,0}$ and $x_{t,1}$ are the $t$-th noise-corrupted observation sample of $H_0$ and $H_1$, respectively, and $T$ denotes the total number of training samples. The observation $x_{t,0}$ and $x_{t,1}$ are obtained from the sensors by periodically observing the desired state $H_0$ and $H_1$ under the practical environment. At each epoch, based on the mini-batch method \cite{o2017introduction}, the whole dataset is divided into $T/B$ batches, where $B$ is the number of batch samples. For the training of $\Phi$ during each batch, the loss function $\mathcal{L}_{\Phi}$ is approximated and averaged on the whole batch samples as 
\begin{equation}\label{eq:approximate_loss_phi}
	\begin{aligned}
		\hat{P}_E^{\text{binary}}=
		\sum_{k=0}^{K}C_{K}^{k}\min&\{\pi_0(\gamma_{\Phi,0})^kp(1-\gamma_{\Phi,0})^{K-k},\\
		&\pi_1(\gamma_{\Phi,1})^kp(1-\gamma_{\Phi,1})^{K-k}\},
	\end{aligned}
\end{equation}
where 
\begin{equation}\label{eq:approximate_gamma_phi}
	\gamma_{\Phi,n}=\sum_{t=1}^{B}G_{\Phi}(x_{t,n}),\ n=0,1,
\end{equation}
is the empirical approximation of $\gamma_{\Phi}(H_n)=\mathbb{E}_{X}[G_{\Phi}(X)|H_n]$ over the whole batch.  The optimal probability controller parameters are obtained as $\Phi^*$ when the maximum number of training epochs is reached.

Once the probability controller parameters $\Phi^*$ are obtained, they can be utilized to train the detector parameters $\Theta$ at the FC. Training the detector $\Theta$ uses the same mini-batch method with the same data set $D_1$. The loss function in \eqref{eq:KL_divergence} is approximated and averaged on the whole batch samples as
\begin{equation}
	\label{eq:approximate_KL_divergence}
	\begin{aligned}
		\hat{D}_{KL}^{\text{binary}}
		=&\sum_{n=0}^{1}\sum_{k=0}^{K}
		\pi_nC_K^k(\gamma_{n}^*)^k(1-\gamma_{n}^*)^{K-k}\\
		&\times\log\frac{\pi_nC_K^k(\gamma_{n}^*)^k(1-\gamma_{n}^*)^{K-k}}{F_{\Theta}(H_n|\bar{u})\sum_{n=0}^{1}\pi_nC_K^k(\gamma_{n}^*)^k(1-\gamma_{n}^*)^{K-k}},
	\end{aligned}
\end{equation}
with $\gamma_n^*=\sum_{t=1}^{B}G_{\Phi^*}(x_{t,n}),\ n=0,1$. Similarly, the optimal detector parameters are obtained as $\Theta^*$ when the maximum number of training epochs is reached.

\section{Simulation results}\label{sec:simulation}
This section presents simulation results to show the superior performance of the proposed method. We consider the observations of the desired state of nature at all sensors are corrupted by the Gaussian noise, i.e.,
\begin{align}
	&f_X|H(\cdot|H_0)\sim\mathcal{N}(-1,\sigma^2),\\
	&f_X|H(\cdot|H_1)\sim\mathcal{N}(1,\sigma^2).
\end{align}
The samples of the observations in both the data and test sets are obtained from sensors by observing the desired state $H$ with prior probabilities $\pi_0=\pi_1=0.5$. 
The signal-to-noise ratio (SNR) is defined as the power ratio of the desired state to the observation noise, i.e., $\text{SNR}=1/\sigma^2$.

\begin{figure}[htbp]
	\centering
	\subfigure[Training loss of probability controller]{
		\includegraphics[width=0.4\textwidth]{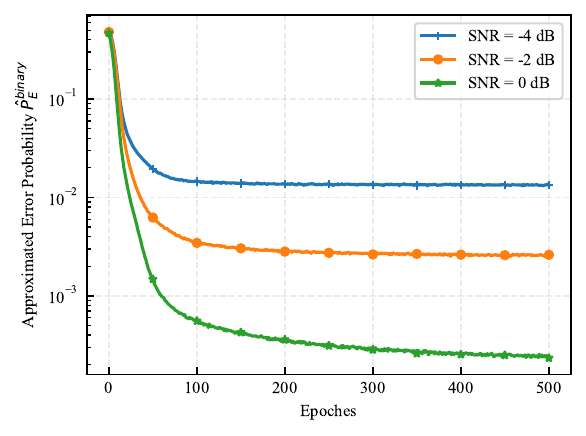}\label{subfig:loss_quantizer} 
	}
	\subfigure[Training loss of detector]{
		\includegraphics[width=0.4\textwidth]{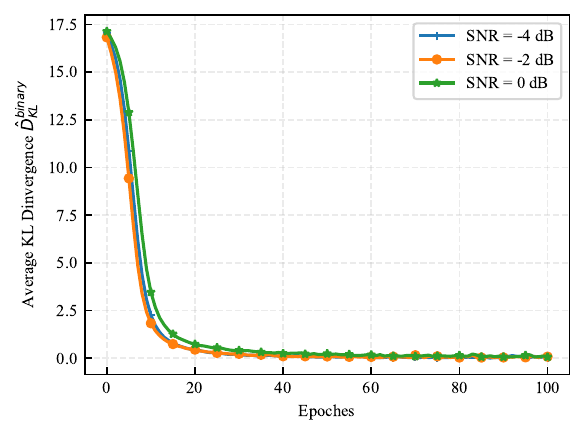} \label{subfig:loss_detector} 
	}
	\caption{Training loss of probability controller and detector.}
	\label{fig:convergence}\vspace{-1em}
\end{figure}

For the probability controller DNN $\Phi$ in the quantizer, we use a multi-layer perceptron (MLP) with the number of hidden layers $L=3$ and the number of neurons in each layer is $M=20$.
For the detector DNN $\Theta$ at the FC, we consider a MLP with the same number of hidden layers and the number of neurons in each layer is $N=30$. The number of sensors used in the training is $K=20$. $\Phi$ is trained with 50, 000 samples and the number of epochs is 500.  Using the trained $\Phi^*$, $\Theta$ is then trained by the same dataset. After training $\Phi$ and $\Theta$ separately, the whole system is tested by a set with 10, 000 samples. The whole training uses the Adam optimizer\cite{o2017introduction} with a constant learning rate $l_r=0.0001$. The implementation of whole simulation is carried out using PyTorch 1.7.0 \cite{o2017introduction}. 

Fig. \ref{fig:convergence} shows the training loss of the proposed probability controller and detector with the number of epoch. It is observed that with the increasing of the training epochs, the loss of the quantizer training quickly decrease and converges to a constant. Since the approximated average probability of detection error, $\hat{P}_{E}^{\text{binary}}$ defined in \eqref{eq:approximate_loss_phi}, is utilized as the training loss, the result implies that smaller true detection error probability can be achieved. Fig. \ref{subfig:loss_detector} illustrates the training loss of the detector with the number of epochs. It is observed that with the increasing of the training epochs, the training loss of the detector quickly decreases and converges to a constant near zero. Since the average KL divergence between the true posterior probability and its DNN approximation is utilized as the training loss for the detector, the result shows that better approximation of the true posterior probability is achieved, which further implies that the better detection performance is obtained by the detector. 

\begin{figure}[htbp]
	\centering
	\includegraphics[width=0.4\textwidth]{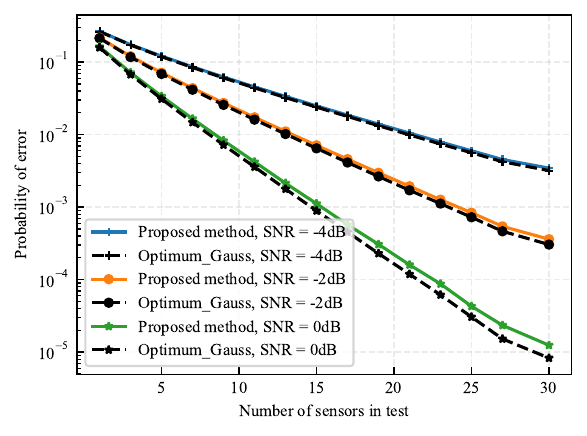}
	\caption{Probability of detection error vs. the number of sensors under different SNRs.}
	\label{fig:detection_performance}\vspace{-1em}
\end{figure}

Fig. \ref{fig:detection_performance} shows the performance of the average detection error for the proposed method under the stable noisy scenario, where the SNRs in the training and test stages are the same. For comparisons, we implement the numerical simulation of the Gaussian approximation method \cite{gul2021fast}, which is optimal for the Gaussian hypothesis testing problems. It is observed that under different SNRs, the detection performance of the proposed method is close to the optimal one and shows exponentially decreasing trend with the number of sensors. However, the proposed method has much lower complexity in the order of $\mathcal{O}(KL(M+N))$ compared to the Gaussian approximation method in the order of $\mathcal{O}(K^2\log K)$.

\section{Conclusion}\label{sec:conclusion}
In this paper, we proposed a model-driven DL-based method for the distributed detection with binary quantization in WSNs. First, we derived the lower bound of the detection error probability based on the MAP decision rule and proved that utilizing an identical quantizer across the sensors is globally optimal in terms of maximizing corresponding Chernoff information. Then, we derived MAPDEP considering an identical binary probabilistic quantizer. Next, we proved the equivalence between using the full quantized data and their average as input to the detector at the FC, and derived the KL divergence to measure the performance of the detector. Finally, we proposed a sequential training method to train the probability controller and detector modules. Simulation results validated the superior performance of the proposed method.

\appendices
\section{Proof of Lemma \ref{lem:glob_opt_identical_quantizer}}\label{append:proof_identical_quantizer}
Based on the expression of $P_E(H,\mathbf{u})$ in \eqref{eq:MAPDEP} and the definition of the corresponding average Chernoff information $C(H,\mathbf{u})=-\frac{1}{K}\log P_E(H,\mathbf{u})$, we have 
\begin{equation}\label{eq:Chernoff_Bound}
	\begin{aligned}
		C(H,\mathbf{u})
		=&-\frac{1}{K}\log\left[\sum_{\mathbf{u}}\min\left\{\pi_0p(\mathbf{u}|H_0),\pi_1p(\mathbf{u}|H_1)\right\}\right]\\
		\leq&-\frac{1}{K}\log\left[\sum_{\mathbf{u}}\left(\pi_0p(\mathbf{u}|H_0)\right)^{\alpha}\left(\pi_1p(\mathbf{u}|H_1)\right)^{1-\alpha}\right]\\
		\leq& -\frac{1}{K}\log\left[\sum_{\mathbf{u}}\left(p(\mathbf{u}|H_0)\right)^{\alpha}\left(p(\mathbf{u}|H_1)\right)^{1-\alpha}\right],
	\end{aligned}
\end{equation}
where the first inequality is due to $\min\{x,y\}\leq x^{\alpha}y^{1-\alpha},\ \forall\ 0\leq\alpha\leq 1, \forall x>0,y>0$ and the achievability of the equality is shown in \cite{cover2006elements}. Therefore, the average Chernoff information bound takes the minimum over $0\leq\alpha\leq 1$, i.e., 
\begin{equation}\label{eq:Chernoff_info_def}
	C(H,\mathbf{u})=-\frac{1}{K}\underset{0\leq \alpha\leq 1}{\min} \log \left[
	\sum_{\mathbf{u}}\left(p(\mathbf{u}|H_{0})\right)^{\alpha}\left(p(\mathbf{u}|H_{1})\right)^{1-\alpha}\right].
\end{equation}
Mathematically, for $\mathbf{u}=[u_1,\cdots,u_K]$ and $n\in\{0,1\},$ we have
\begin{equation}\label{eq: p(mathbf{u}|H) 1}
	p(\mathbf{u}|H_n)=\prod_{k=1}^{K}p(u_k|H_n).
\end{equation}
By subsisting \eqref{eq: p(mathbf{u}|H) 1} into \eqref{eq:Chernoff_info_def}, the Chernoff information is rewritten as  
\begin{equation}\label{eq:Chernoff_Bound_New}
	\begin{aligned}
		&C(H,\mathbf{u})\\
		=& -\frac{1}{K}\underset{0\leq \alpha\leq 1}{\min} \log\left[
		\sum_{\mathbf{u}}p(\mathbf{u}|H_{0})^{\alpha}p(\mathbf{u}|H_{1})^{1-\alpha}\right]\\
		=& -\frac{1}{K}\underset{0\leq \alpha\leq 1}{\min} \log\left[
		\sum_{u_1=0}^{L-1}\sum_{u_2=0}^{L-1}\cdots\sum_{u_K=0}^{L-1}\left(\prod_{k=1}^{K}p(u_k|H_{0})^{\alpha}\right.\right.\\
		&\left.\left.\times p(u_k|H_{1})^{1-\alpha}\right)\right]\\
		=& -\frac{1}{K}\underset{0\leq \alpha\leq 1}{\min} \log\left[
		\prod_{k=1}^{K}\left(\sum_{u_k=0}^{L-1}
		p(u_k|H_{0})^{\alpha}p(u_k|H_{1})^{1-\alpha}\right)\right]\\
		=& -\frac{1}{K}\underset{0\leq \alpha\leq 1}{\min} \sum_{k=1}^{K}\log\left[\sum_{u_k=0}^{L-1}
		p(u_k|H_{0})^{\alpha}p(u_k|H_{1})^{1-\alpha}\right],
\end{aligned}\end{equation}
and can be upper bounded by
\begin{equation}\label{eq:Chernoff_Bound_New1}
	\begin{aligned}
		&C(H,\mathbf{u})\\
		=& -\frac{1}{K}\underset{0\leq \alpha\leq 1}{\min} \sum_{k=1}^{K}\log\left[\sum_{u_k=0}^{L-1}
		p(u_k|H_{0})^{\alpha}p(u_k|H_{1})^{1-\alpha}\right]\\
		\leq& -\frac{1}{K}\sum_{k=1}^{K}\underset{0\leq \alpha_k\leq 1}{\min} \log\left[\sum_{u_k=0}^{L-1}
		p(u_k|H_{0})^{\alpha_k}p(u_k|H_{1})^{1-\alpha_k}\right]\\
		=& -\frac{1}{K}\sum_{k=1}^{K} \log\left[\sum_{u_k=0}^{L-1}
		p(u_k|H_{0})^{\alpha_k^*}p(u_k|H_{1})^{1-\alpha_k^*}\right],
\end{aligned}\end{equation}
where
\begin{equation}
	\alpha_k^*=\arg\mathop{\mathrm{min}}
	\limits_{0\leq\alpha_k \leq1} \log\left[\sum_{u_k=0}^{L-1}
	p(u_k|H_{0})^{\alpha_k}p(u_k|H_{1})^{1-\alpha_k}\right].
\end{equation}
The equality in \eqref{eq:Chernoff_Bound_New1} holds if and only if
$\alpha_1^*=\alpha_2^*=\cdots=\alpha_K^*$, which means that the quantized information $u_1,u_2,\ldots,u_K$ should be conditionally i.i.d. with given $H$, i.e., $p(\mathbf{u}|H_n)=\prod_{k=1}^{K}p(u_k|H_n)$ and $p(u_1=l|H_n)=p(u_2=l|H_n)=\cdots=p(u_K=l|H_n), l\in\{0,1,\ldots,L-1\}, n\in\{0,1\}$. This implies that identical quantizer are deployed at all sensors. In this case, the Chernoff information is rewritten as  
\begin{equation}\label{eq:Chernoff_Info_New}
	C(H,\mathbf{u})=-\underset{0\leq \alpha\leq 1}{\min} \log\left[\sum_{u=0}^{L-1}p(u|H_{0})^{\alpha}p(u|H_{1})^{1-\alpha}\right].
\end{equation}

\section{Proof of Proposition \ref{prop:MAPDEP_binary_quantize}}
\label{append:proof_MAPDEP_binary_quantize}
Define $\mathcal{U}=\{\mathbf{u}=[u_1,\cdots,u_K]^T|\ u_k=0\ \text{or}\ 1,\ k =1,2,\dots,K\}=\{0,1\}^K$ and the sequence of sets $\{\mathcal{U}_k\}_{k=0}^K$ with  $\mathcal{U}_k=\{\mathbf{u}\in\mathcal{U}|\frac{1}{K}\sum_{i=1}^K u_i=k/K\}$. If the binary quantized data $u_1,u_2,\ldots,u_K$ are conditionally i.i.d. with given $H$, we have that any $\mathbf{u}$ belonging to set $\mathcal{U}_k$ have the identical conditional probability with given $H$, i.e.,
\begin{equation}\label{eq:bernoulli_distribution}
	\begin{aligned}		 
		p(\mathbf{u}|H)
		=&(p(u=1|H))^k(p(u=0|H))^{K-k}\\
		=&(\gamma(H))^k(1-\gamma(H))^{K-k},
	\end{aligned}
\end{equation}   
$\forall \mathbf{u}\in \mathcal{U}_k, H\in\{H_0,H_1\}$, where $\gamma(H)\triangleq p(u=1|H)$ denotes the conditional probability of any quantized data being `1' with given $H$. By using \eqref{eq:bernoulli_distribution} and $|\mathcal{U}_k|=C_K^k$, the minimum achievable MAPDEP of the binary quantized based detection is given as 
\begin{equation}
	\label{eq: minimum error probability of the binary-quantization-based detection 1}
	\begin{aligned}
		P_{E}(H,\mathbf{u})
		=&\sum_{\mathbf{u}}\min\left\{\pi_0p(\mathbf{u}|H_0),\pi_1p(\mathbf{u}|H_1)\right\}\\
		=&\sum_{k=0}^{K}\sum_{\mathbf{u}\in\mathcal{U}_k}\min\left\{\pi_0p(\mathbf{u}|H_0),\pi_1p(\mathbf{u}|H_1)\right\}\\
		=&\sum_{k=0}^{K}C_{K}^{k}\min\{\pi_0(\gamma(H_0))^kp(1-\gamma(H_0))^{K-k},\\
		&\pi_1(\gamma(H_1))^kp(1-\gamma(H_1))^{K-k}\}.
	\end{aligned}
\end{equation} 
This completes the proof of Proposition \ref{prop:MAPDEP_binary_quantize}.

\section{Proof of Theorem \ref{th:opt_mean_fusion}}\label{append:proof_mean_fusion}
Based on \eqref{eq:bernoulli_distribution} and using $|\mathcal{U}_k|=C_K^k$, the conditional probability distribution for the average  $\bar{u}$ of all quantized data with given $\theta$ is computed as
\begin{equation}\label{eq: bernoulli_binomial_distribution}
	\begin{aligned}
		p\left(\bar{u}=\frac{k}{K}\bigg|H\right)
		=&p(\mathbf{u}\in\mathcal{U}_k|H)\\
		=&C_K^k(p(u=1|H))^k(p(u=0|H))^{K-k}\\
		=&C_K^k(\gamma(H))^k(1-\gamma(H))^{K-k},
	\end{aligned}
\end{equation}
$k=0,1,\cdots,K, H\in\{H_0,H_1\}$. Based on \eqref{eq: bernoulli_binomial_distribution}, the minimum detection error probability by using $\bar{u}$ for detection of $H$ is given as
\begin{equation}
	\begin{aligned}
		&P_E(H,\bar{u})\\
		=&\sum_{\bar{u}}\min\left\{\pi_0p(\bar{u}|H_0),\pi_1p(\bar{u}|H_1)\right\}\\
		=&\sum_{k=0}^{K}\min\left\{\pi_0p\left(\bar{u}=\frac{k}{K}\bigg|H_0\right),\pi_1p\left(\bar{u}=\frac{k}{K}\bigg|H_1\right)\right\}\\
		=&\sum_{k=0}^{K}C_{K}^{k}\min\{\pi_0(\gamma(H_0))^kp(1-\gamma(H_0))^{K-k},\\
		&\pi_1(\gamma(H_1))^kp(1-\gamma(H_1))^{K-k}\}.
	\end{aligned}
\end{equation}
This shows that $P_E(H,\mathbf{u})=P_E(H,\bar{u})$, which completes the proof of Theorem \ref{th:opt_mean_fusion}.

\section{Proof of Proposition \ref{prop:KL_diverngence}}
\label{append:proof_KL_divergence}
The average KL divergence between the posterior probability $P(\bar{u}|H)$ and its approximation $F_{\Theta}(H|\bar{u})$ is given as
\begin{equation}\label{eq:KL_divergence_proof}
	\begin{aligned} &\mathbb{E}_{\bar{u}}\left[D_{KL}(P(H|\bar{u})|F_{\Theta}(H|\bar{u}))\right]\\
		=&\mathbb{E}_{\bar{u}}\left[\mathbb{E}_{P(H|\bar{u})}\left[\log\frac{P(H|\bar{u})}{F_{\Theta}(H|\bar{u})}\right]\right]\\
		=&\sum_{n=0}^{1}\sum_{k=0}^{K}p\left(H_n,\bar{u}=\frac{k}{K}\right)\log\frac{p\left(H_n\left|\bar{u}=\frac{k}{K}\right.\right)}{F_{\Theta}(H_n|\bar{u})}.
	\end{aligned}
\end{equation}
Based on the Bayesian rules, we have that
\begin{align}
	\label{eq:Bayesian_Proof_1}
	p\left(H_n,\bar{u}=\frac{k}{K}\right)
	=&\ \pi_np\left(\left.\bar{u}=\frac{k}{K}\right|H_n\right),\\ 
	\label{eq:Bayesian_Proof_2}
	p\left(H_n\left|\bar{u}=\frac{k}{K}\right.\right)
	=&\ \frac{p\left(H_n,\bar{u}=\frac{k}{K}\right)}{p\left(\bar{u}=\frac{k}{K}\right)}\nonumber\\
	=&\frac{\pi_np\left(\left.\bar{u}=\frac{k}{K}\right|H_n\right)}{\sum_{n=0}^{1}\pi_np\left(\left.\bar{u}=\frac{k}{K}\right|H_n\right)}.
\end{align}
By subsisting \eqref{eq:Bayesian_Proof_1} and \eqref{eq:Bayesian_Proof_2} into \eqref{eq:KL_divergence_proof}, the average KL divergence with the given parameters $\Phi$ and $\Theta$ is rewritten as
\begin{equation}\label{eq:KL_divergence_proof_new}
	\begin{aligned} 
		&D_{KL}^{Binary}(\Phi,\Theta)\\
		=& 			    
		\mathbb{E}_{\bar{u}}\left[D_{KL}(P(H|\bar{u})|F_{\Theta}(H|\bar{u}))|\Phi,\Theta\right]\\
		=& \sum_{n=0}^{1}\sum_{k=0}^{K}
		\pi_np\left(\left.\bar{u}=\frac{k}{K}\right|H_n,\Phi\right)\\
		&\times\log\frac{\pi_np\left(\left.\bar{u}=\frac{k}{K}\right|H_n,\Phi\right)}{F_{\Theta}(H_n|\bar{u})\sum_{n=0}^{1}\pi_np\left(\left.\bar{u}=\frac{k}{K}\right|H_n,\Phi\right)}\\
		=& \sum_{n=0}^{1}\sum_{k=0}^{K}
		\pi_np_{\Phi}^{k}(H_n)
		\log\frac{\pi_np_{\Phi}^{k}(H_n)}{F_{\Theta}(H_n|\bar{u})\sum_{n=0}^{1}\pi_np_{\Phi}^{k}(H_n)},
	\end{aligned}
\end{equation}
where $p_{\Phi}^{k}(H_n)=p\left(\left.\bar{u}=\frac{k}{K}\right|H_n,\Phi\right)=C_K^k(\gamma_{\Phi}(H_n))^k(1-\gamma_{\Phi}(H_n))^{K-k}$.

\bibliographystyle{IEEEtran}
\bibliography{reference}
	
\end{document}